\font\bigbold=cmbx12   \font\ninerm=cmr9      \font\eightrm=cmr8
\font\sixrm=cmr6       \font\fiverm=cmr5      \font\ninebf=cmbx9
\font\eightbf=cmbx8    \font\sixbf=cmbx6      \font\fivebf=cmbx5
\font\ninei=cmmi9      \skewchar\ninei='177   \font\eighti=cmmi8
\skewchar\eighti='177  \font\sixi=cmmi6       \skewchar\sixi='177
\font\fivei=cmmi5      \font\ninesy=cmsy9     \skewchar\ninesy='60
\font\eightsy=cmsy8    \skewchar\eightsy='60  \font\sixsy=cmsy6
\skewchar\sixsy='60    \font\fivesy=cmsy5     \font\nineit=cmti9
\font\eightit=cmti8    \font\ninesl=cmsl9     \font\eightsl=cmsl8
\font\ninett=cmtt9     \font\eighttt=cmtt8    \font\tenfrak=eufm10
\font\ninefrak=eufm9   \font\eightfrak=eufm8  \font\sevenfrak=eufm7
\font\fivefrak=eufm5   \font\tenbb=msbm10     \font\ninebb=msbm9
\font\eightbb=msbm8    \font\sevenbb=msbm7    \font\fivebb=msbm5
\font\tenssf=cmss10    \font\ninessf=cmss9    \font\eightssf=cmss8
\font\tensmc=cmcsc10

\newfam\bbfam   \textfont\bbfam=\tenbb \scriptfont\bbfam=\sevenbb
\scriptscriptfont\bbfam=\fivebb  \def\Bbb{\fam\bbfam}
\newfam\frakfam  \textfont\frakfam=\tenfrak \scriptfont\frakfam=%
\sevenfrak \scriptscriptfont\frakfam=\fivefrak  \def\frak{\fam\frakfam}
\newfam\ssffam  \textfont\ssffam=\tenssf \scriptfont\ssffam=\ninessf
\scriptscriptfont\ssffam=\eightssf  
\def\smc{\tensmc}

\def\eightpoint{\textfont0=\eightrm \scriptfont0=\sixrm
\scriptscriptfont0=\fiverm  \def\rm{\fam0\eightrm}%
\textfont1=\eighti \scriptfont1=\sixi \scriptscriptfont1=\fivei
\def\oldstyle{\fam1\eighti}\textfont2=\eightsy
\scriptfont2=\sixsy \scriptscriptfont2=\fivesy
\textfont\itfam=\eightit         \def\it{\fam\itfam\eightit}%
\textfont\slfam=\eightsl         \def\sl{\fam\slfam\eightsl}%
\textfont\ttfam=\eighttt         \def\tt{\fam\ttfam\eighttt}%
\textfont\frakfam=\eightfrak     \def\frak{\fam\frakfam\eightfrak}%
\textfont\bbfam=\eightbb         \def\Bbb{\fam\bbfam\eightbb}%
\textfont\bffam=\eightbf         \scriptfont\bffam=\sixbf
\scriptscriptfont\bffam=\fivebf  \def\bf{\fam\bffam\eightbf}%
\abovedisplayskip=9pt plus 2pt minus 6pt   \belowdisplayskip=%
\abovedisplayskip  \abovedisplayshortskip=0pt plus 2pt
\belowdisplayshortskip=5pt plus2pt minus 3pt  \smallskipamount=%
2pt plus 1pt minus 1pt  \medskipamount=4pt plus 2pt minus 2pt
\bigskipamount=9pt plus4pt minus 4pt  \setbox\strutbox=%
\hbox{\vrule height 7pt depth 2pt width 0pt}%
\normalbaselineskip=9pt \normalbaselines \rm}

\def\ninepoint{\textfont0=\ninerm \scriptfont0=\sixrm
\scriptscriptfont0=\fiverm  \def\rm{\fam0\ninerm}\textfont1=\ninei
\scriptfont1=\sixi \scriptscriptfont1=\fivei \def\oldstyle%
{\fam1\ninei}\textfont2=\ninesy \scriptfont2=\sixsy
\scriptscriptfont2=\fivesy
\textfont\itfam=\nineit          \def\it{\fam\itfam\nineit}%
\textfont\slfam=\ninesl          \def\sl{\fam\slfam\ninesl}%
\textfont\ttfam=\ninett          \def\tt{\fam\ttfam\ninett}%
\textfont\frakfam=\ninefrak      \def\frak{\fam\frakfam\ninefrak}%
\textfont\bbfam=\ninebb          \def\Bbb{\fam\bbfam\ninebb}%
\textfont\bffam=\ninebf          \scriptfont\bffam=\sixbf
\scriptscriptfont\bffam=\fivebf  \def\bf{\fam\bffam\ninebf}%
\abovedisplayskip=10pt plus 2pt minus 6pt \belowdisplayskip=%
\abovedisplayskip  \abovedisplayshortskip=0pt plus 2pt
\belowdisplayshortskip=5pt plus2pt minus 3pt  \smallskipamount=%
2pt plus 1pt minus 1pt  \medskipamount=4pt plus 2pt minus 2pt
\bigskipamount=10pt plus4pt minus 4pt  \setbox\strutbox=%
\hbox{\vrule height 7pt depth 2pt width 0pt}%
\normalbaselineskip=10pt \normalbaselines \rm}

\global\newcount\secno \global\secno=0 \global\newcount\meqno
\global\meqno=1 \global\newcount\appno \global\appno=0
\newwrite\eqmac \def\romappno{\ifcase\appno\or A\or B\or C\or D\or
E\or F\or G\or H\or I\or J\or K\or L\or M\or N\or O\or P\or Q\or
R\or S\or T\or U\or V\or W\or X\or Y\or Z\fi}
\def\eqn#1{ \ifnum\secno>0 \eqno(\the\secno.\the\meqno)
\xdef#1{\the\secno.\the\meqno} \else\ifnum\appno>0
\eqno({\rm\romappno}.\the\meqno)\xdef#1{{\rm\romappno}.\the\meqno}
\else \eqno(\the\meqno)\xdef#1{\the\meqno} \fi \fi
\global\advance\meqno by1 }

\global\newcount\refno \global\refno=1 \newwrite\reffile
\newwrite\refmac \newlinechar=`\^^J \def\ref#1#2%
{\the\refno\nref#1{#2}} \def\nref#1#2{\xdef#1{\the\refno}
\ifnum\refno=1\immediate\openout\reffile=refs.tmp\fi
\immediate\write\reffile{\noexpand\item{[\noexpand#1]\ }#2\noexpand%
\nobreak.} \immediate\write\refmac{\def\noexpand#1{\the\refno}}
\global\advance\refno by1} \def\semi{;\hfil\noexpand\break ^^J}
\def\nl{\hfil\noexpand\break ^^J} \def\refn#1#2{\nref#1{#2}}

\def\ann#1#2#3{{\it Ann.\ Phys.}\ {\bf {#1}} ({#2}) #3}

\def\jmp#1#2#3{{\it J.\ Math.\ Phys.}\ {\bf {#1}} ({#2}) #3}
\def\jpA#1#2#3{{\it J.\ Phys.}\ {\bf A{#1}} ({#2}) #3}

\def\plA#1#2#3{{\it Phys.\ Lett.}\ {\bf {#1}A} ({#2}) #3}

\def\prA#1#2#3{{\it Phys.\ Rev.}\ {\bf A{#1}} ({#2}) #3}

\def\prD#1#2#3{{\it Phys.\ Rev.}\ {\bf D{#1}} ({#2}) #3}
\def\prl#1#2#3{{\it Phys.\ Rev.\ Lett.}\ {\bf #1} ({#2}) #3}

\newif\iftitlepage \titlepagetrue \newtoks\titlepagefoot
\titlepagefoot={\hfil} \newtoks\otherpagesfoot \otherpagesfoot=%
{\hfil\tenrm\folio\hfil} \footline={\iftitlepage\the\titlepagefoot%
\global\titlepagefalse \else\the\otherpagesfoot\fi}

\def\abstract#1{{\parindent=30pt\narrower\noindent\ninepoint\openup
2pt #1\par}}

\newcount\notenumber\notenumber=1 \def\note#1
{\unskip\footnote{$^{\the\notenumber}$} {\eightpoint\openup 1pt #1}
\global\advance\notenumber by 1}
%

\def\today{\ifcase\month\or January\or February\or March\or
April\or May\or June\or July\or August\or September\or October\or
November\or December\fi \space\number\day, \number\year}

\def\pagewidth#1{\hsize #1}  \def\hcorrection#1{\advance\hoffset by #1}
\def\pageheight#1{\vsize #1} \def\vcorrection#1{\advance\voffset by #1}

\pageheight{23cm}
\pagewidth{15.7cm}
\hcorrection{-1mm}
\magnification= \magstep1
\parskip=5pt plus 1pt minus 1pt
\tolerance 8000
\def\bsk{\baselineskip= 14.5pt plus 1pt minus 1pt}
\bsk

\font\extra=cmss10 scaled \magstep0  \setbox1 = \hbox{{{\extra R}}}
\setbox2 = \hbox{{{\extra I}}}       \setbox3 = \hbox{{{\extra C}}}
\setbox4 = \hbox{{{\extra Z}}}       \setbox5 = \hbox{{{\extra N}}}




                                                        
\def\frac#1#2{{#1\over#2}}

\def\pmb#1{\setbox0=\hbox{$#1$} \kern-.025em\copy0\kern-\wd0
    \kern.05em\copy0\kern-\wd0 \kern-.025em\raise.0433em\box0 }

\def\ve{\vfill\eject}

\def\Z{{\Zed}}
\def\R{{\Real}}
\def\C{{\Complex}}
\def\({\left(}
\def\){\right)}

\def\HB{\hbox} \def\SB{\setbox1\HB}  
 
\def\XB{\wd1}  \def\CB{\copy1} \def\RB#1{\raise#1\CB}
  \def\CC{\copy2} \def\RC#1{\raise#1\CC}

        \def\yph{\vphantom}       
\def\x{\hskip}             \def\y{\vskip}
\def\xex#1{\x #1 ex}       \def\yex#1{\y #1 ex}
\def\xpt#1{\x #1 truept}

\def\hpt{\xpt}    
    \def\vph{\yph}

\def\bit{\hskip 0.15ex}      \def\biT{\hskip -0.15ex}
\def\bitt{\hskip 0.30ex}     \def\biTT{\hskip -0.30ex}
\def\bittt{\hskip 0.45ex}

   \def\newline{\hfill\break}

\def\sectit#1{\bigskip\noindent{\bf #1}\smallskip}
\def\quotes#1{`#1'}        
     
\def\emp#1{{\it #1}}

\def\eg{{\it e.g.,\ }}     
\def\ie{{\it i.e.,\ }}

\def\cf{cf.\ }             
          

\def\Bf#1{\HB{$\SB{$#1$}     
          \def\P{\CB\x-\XB\RB{.05ex}}\P\x-\XB\x.07ex\P$}}

\def\m#1{$             #1          $}  
\def\mm#1{$    \,      #1  \,      $}  
\def\mmm#1{$   \,\,    #1  \,\,    $}  

\def\f#1#2{{#1\over#2}}     
\def\ff#1#2{\raise.5pt\hbox{\eightpoint${\displaystyle\f{#1}{#2}}$}}






\def\arrow#1{\buildrel #1 \over \longmapsto}

\def\ordo#1{ {\cal O} \biTT \( #1 \) }

\def\lt{\left}    \def\({\lt(}    \def\[{\lt[}    \def\<{\lt\langle}
\def\rt{\right}   \def\){\rt)}    \def\]{\rt]}    \def\>{\rt\rangle}

\def\impl{\Rightarrow}    

\def\homeo{\cong}

\def\text#1{{   \rm     #1    }}     
\def\textt#1{{  \rm     #1 \, }}     
\def\texttt#1{{ \rm     #1 \  }}     
     
\def\d{\text{d}}
\def\arccot{\textt{arccot}}

\def\eps{\varepsilon}  \def\phy{\varphi}  \def\teta{\vartheta}

\def\h{\hbar}
\def\R{{\rm I \hskip-0.47ex R}}
\def\R{{\Bbb R}}    \def\Z{{\Bbb Z}}  \def\C{{\Bbb C}}
\def\co{\, ,}   \def\pe{\, .}   

\def\aR{\alpha_{\rm R}}  \def\aI{\alpha_{\rm I}}
\def\bR{\beta_{ \rm R}}  \def\bI{ \beta_{\rm I}}

\def\PL{^{(+)}}  \def\MI{^{(-)}}  \def\PM{^{(\pm)}}  \def\MP{^{(\mp)}}

\def\P{{\cal P}}
\def\qQ{{\cal Q}}
\def\qR{{\cal R}}

\def\T{{\cal T}}
\def\qF{{\cal F}}
\def\qFP{{\cal F}_{{\cal P}}}
\def\qFT{{{\cal F}_{{\cal T}}}}
\def\qFPT{{{\cal F}_{{\cal PT}}}}
\def\qFSD{{\cal F}_{\rm SD}}
\def\qV{{\cal V}}
\def\qW{{\cal W}}
\def\PT{{\cal PT}}

\def\Ll{L_{\rm left}}  \def\Lr{L_{\rm right}}
\def\Lr{L_{\matrix{ {\rm left} \cr {\rm \right}}}}
\def\Ll{L_1}           \def\Lr{L_2}
\def\Llr{L_{0 \xex{-.8} \SB{$\scriptstyle l$} \RB{-1.6ex}}}
\def\Llr{L_{1, \bit 2}}

\def\QPIL{\quotes{point interaction on a line}}
\def\qHam{H}
  
\def\II{I_2}
\def\IIII{I_4}
\def\Psix{\Phi}
\def\pmeps{\varepsilon}

\def\E#1#2{$$ #2 \eqn{#1} $$}
\def\_{^{}_}


{
\refn\Ann
{T. Cheon, T. F\"ul\"op and I. Tsutsui,
\ann{294}{2001}{1}}

\refn\Seba
{P. \v{S}eba,
{\it Czech. J. Phys.} {\bf 36} (1986) 667}

\refn\AGHH
{S. Albeverio, F. Gesztesy, R. H{\o}egh-Krohn and H. Holden,
\lq\lq Solvable Models in Quantum Mechanics\rq\rq,
Springer, New York, 1988}

\refn\CFG
{M. Carreau, E. Farhi and S. Gutmann,
\prD{42}{1990}{1194}}

\refn\CH
{P.R. Chernoff and R.J. Hughes,
{\it J. Funct. Anal.} {\bf 111} (1993) 97}

\refn\ABD
{S. Albeverio, Z. Brze\'{z}niak and L. Dabrowski,
{\it J. Funct. Anal.} {\bf 130} (1995) 220}

\refn\LuzCheng
{M.G.E. da Luz and B.K. Cheng,
\prA{51}{1995}{1811}}

\refn\FT
{T. F\"{u}l\"{o}p and I. Tsutsui, \plA{264}{2000}{366}}

\refn\Warburton
{R.J. Warburton, C. Sch\"aflein, D. Haft, F. Bickel, A. Lorke, K. Karrai, J.M.
Garcia, W. Scoenfeld and P.M. Petroff, {\it Nature} {\bf 405} (2000) 926}

\refn\Lorke
{A. Lorke, R.J. Luyken, A.O. Govorov, J.P. Kotthaus, J.M. Garcia and P.M.
Petroff, \prl{84}{2000}{2223}}

\refn\Nature
{A. Fuhrer, S. L{\"u}sher, T. Ihn, T. Heinzel, K. Ensslin, W. Wegscheider
and M. Bichler, {\it Nature} {\bf 413} (2001) 822}

\refn\JMP
{I. Tsutsui, T. F\"ul\"op and T. Cheon, \jmp{42}{2001}{5687}}

\refn\Car
{M. Carreau, \jpA{26}{1993}{427}}

\refn\RT
{J.M. Rom\'{a}n and R. Tarrach, \jpA{29}{1996}{6073}}

\refn\Kurasov
{P. Kurasov, {\it Journ. Math. Anal. Appl.} {\bf 201} (1996) 297}

\refn\SMMC
{T. Cheon and T. Shigehara, {\it Phys. Lett.} {\bf 243A} (1998) 111}

\refn\AN
{S. Albeverio and L. Nizhnik, {\it Ukrainian Math. J.} {\bf 52} (2001) 644}

\refn\ENZ
{P. Exner, H. Neidhardt, V. Zagrebnov,
{\it Commun. Math. Phys.} {\bf 224} (2001) 593}

\refn\Wall
{T. F\"ul\"op, T. Cheon and I. Tsutsui, \prA{66}{2002}{052102}}

\refn\RS
{M. Reed and B. Simon,
\lq\lq Methods of Modern Mathematical Physics II, Fourier analysis,
self-adjointness\rq\rq, Academic Press, New York, 1975}

\refn\Junker
{G. Junker, \lq\lq Supersymmetric Methods in Quantum and Statistical
Physics\rq\rq, Springer, Berlin, 1996}

\refn\UTone
{T. Uchino and I. Tsutsui, {\it Nucl. Phys.} {\bf B662} (2003) 447}

\refn\Kobe
{T. Nagasawa, M. Sakamoto and K. Takenaga,
{\it Phys. Lett.} {\bf B562} (2003) 358}

\refn\Schulman{L.S. Schulman, 
\lq\lq Techniques and Applications of Path Integration\rq\rq, John Wiley \&
Sons, New York, 1981}

\refn\duallet
{I. Tsutsui, T. F\"ul\"op and T. Cheon,
{\it J.\ Phys.\ Soc.\ Japan} {\bf 69} (2000) 3473}

\refn\UTtwo
{T. Uchino and I. Tsutsui,
{\it J. Phys. A}: Math. Gen. {\bf 36} (2003) 6821}

\refn\NPT
{Y. Nakamura, Yu.A. Pashkin and J.S. Tsai,
{\it Nature} {\bf 398} (1999) 786}

}



\pageheight{23cm}
\pagewidth{14.8cm}
\hcorrection{0mm}
\magnification= \magstep1
\def\bsk{%
\baselineskip= 16.8pt plus 1pt minus 1pt}
\parskip=5pt plus 1pt minus 1pt
\tolerance 6000



\null
\vskip 42pt

\centerline{\bigbold Spectral Properties on a
Circle with a Singularity}

\vskip 35pt

\centerline{
{\smc  
Tam\'{a}s F\"{u}l\"{o}p\footnote{${}^\dagger$}
{\eightpoint E-mail:\quad fulopt@post.kek.jp},
}
{\smc
Izumi Tsutsui\footnote{${}^\ddagger$}
{\eightpoint E-mail:\quad izumi.tsutsui@kek.jp,
http://research.kek.jp/people/itsutsui/} 
}}

\vskip 3pt
{
\baselineskip=13pt
\centerline{\it Institute of Particle and Nuclear Studies}
\centerline{\it High Energy Accelerator Research Organization (KEK)}
\centerline{\it Tsukuba 305-0801, Japan}
}
\vskip 7pt
\centerline{\rm and}
\vskip 3pt

\vskip 10pt

\centerline{\smc
Taksu Cheon\footnote{${}^*$}
{\eightpoint E-mail:\quad cheon@mech.kochi-tech.ac.jp,
http://www.mech.kochi-tech.ac.jp/cheon/} }

\vskip 3pt
{
\baselineskip=13pt
\centerline{\it Laboratory of Physics}
\centerline{\it Kochi University of Technology}
\centerline{\it Tosa Yamada, Kochi 782-8502, Japan}
}
\yex{6} \centerline{(Received \xex{23}} \yex{9}

\abstract{%
{\bf Abstract.} \quad We investigate the spectral and
symmetry properties of a quantum particle moving on a circle with
a pointlike singularity (or point interaction). We find that, within
the $U(2)$ family of the quantum mechanically allowed distinct
singularities, a $U(1)$ equivalence (of duality-type) exists, and
accordingly the space of distinct spectra is $U(1) \times [ SU(2)
/ U(1) ]$, topologically a filled torus. We explore the relationship
of special subfamilies of
the $U(2)$ family to corresponding symmetries, and identify the
singularities that admit an $N = 2$ supersymmetry.  Subfamilies
that are distinguished in the spectral properties or the WKB
exactness are also pointed out.  The spectral and symmetry properties
are also studied in the context of the circle with two singularities,
which provides a useful scheme to discuss the symmetry properties
on a general basis.}

\yex{1}

KEYWORDS: point interaction, point singularity, quantum mechanics, circle,
spectrum, symmetries, supersymmetry, WKB exactness

\vfill\eject


\pageheight{23cm}
\pagewidth{15.7cm}
\hcorrection{-1mm}
\magnification= \magstep1
\def\bsk{%
\baselineskip= 16.8pt plus 1pt minus 1pt}
\parskip=5pt plus 1pt minus 1pt
\tolerance 8000
\bsk
\voffset 1cm

\sectit{1. Introduction}

Systems with point singularities ({\it i.e.,} contact interactions
or reflecting boundaries) provide an important class of solvable
models in quantum mechanics, allowing for useful applications in
various fields in physics, such as in nuclear or condensed matter
physics.  In spite of being relatively simple, these systems exhibit
a variety of exotic features, including renormalization, Landau
poles, anomalous symmetry breaking, duality, supersymmetry, and
spectral anholonomy [\Ann]. In parallel, the recent developments
of nanotechnology have enabled us to manufacture nanoscale quantum
devices, some of which find an appropriate description in terms of
contact interactions.  The system of a line  with a point singularity
can be realized by a quantum well in one dimension and provides
the simplest example for such systems.  The line system with a
singularity has lately been studied intensively, physical/mathematical
aspects of the system are well-known by now [\Seba--\FT].

In the manufacturing process, one may connect the two ends of the
line, or just insets an antidot in a larger quantum dot to make a
loop.  The resultant circle system with point singularities is
expected to be quite similar to the line system in many respects.
Indeed, the two systems share the same four-parameter $U(2)$ family
for the family of possible point singularities and the same boundary
condition induced by the singularity [\FT].  However, one crucial
difference between them is the different {\it topology}, that is,
the circle system obviously allows a circular flow of the probability
current while the line system does not.  Moreover, the circle system
can trap a magnetic flux inside the loop, and hence is sensitive
to the applied magnetic field responsible for Aharonov-Bohm-type effects.
The actual construction of a quantum ring, as a realization of the
circle system, has been done recently [\Warburton--\Nature], and
the effect of a magnetic flux on the energy spectrum has been
measured experimentally.

In this paper, we explore the spectral and symmetry properties of
the circle system with one singularity taking all possible types
of quantum singularities into consideration.  One of our aims is
to specify the spectral space of the  system, which is important
to determine the range of the energy spectra  that we can obtain
by possibly tuning the parameters of the singularity. We shall
learn that the spectral space is of three dimensions and given by
$\Sigma_{\rm circle} \homeo S^1 \times D^2$.  This may be contrasted
to the spectral space of the line system $\Sigma_{\rm line} \homeo
( S^1 \times S^1 ) / \Z_2$ which is two dimensional [\JMP], and
this shows that the circle system can accommodate more variety in
the spectrum than the line system.  Behind this lies the difference
in the (generalized) symmetry structures of the two systems, and
we shall see that this difference also leads to the distinct features
in the invariant subfamilies of the $U(2)$ defined from the
symmetries.  We also discuss other subfamilies that are distinguished
in the spectral properties or the WKB exactness.

The plan of this paper is as follows.  After the Introduction, we
first present the spectral space of the circle system in sect.2 by
inspecting the spectral conditions.  Then we discuss in sect.3 the
spectral preserving  symmetries and related invariant subfamilies
and, in sect.4, the possibility of supersymmetry on the circle.
Other subfamilies which are distinguished in the spectral properties
and the WKB exactness will be mentioned in sect.5. In sect.6, we
provide a generalized framework accommodating two singularities,
where the symmetries and the spectral properties obtained in the
earlier sections can be confirmed in a more transparent manner.
Finally, sect.7 is devoted to the summary and discussions.   Appendix
A provides the detailed discussion for the determination of the
spectral space $\Sigma_{\rm circle}$, while Appendix B supplements
our argument on the scale invariant subfamily in sect.5.

\ve
\secno=2 \meqno=1

\sectit{2. Spectral space and degeneracy}

Before analyzing the spectral properties of the circle system with a point
singularity, we first recall how the system is defined in quantum
mechanics [\FT].  The crucial ingredient in the definition lies in the
fact that the effect of the singularity is expressed
through a certain boundary condition for the wave functions admitted quantum
mechanically.  Namely, the possible form of the boundary condition is determined by
demanding that, under the presence of the singularity, the Hamiltonian
\mm{\qHam = -
\f{\h^2}{2m}
\f{\d^2}{\d x^2}} of the system be self-adjoint on the space of the wave
functions 
obeying the condition.
In terms of a unitary matrix
\mm{U
\in U(2)}, called the {\it characteristic matrix}, 
the boundary condition that meets the demand is found to be
    $$
    \hpt{0.45} (U - \II) \Psi + i L_0 \, (U + \II) \Psi' = 0 \co \hpt{30}
    \Psi  := \pmatrix{ \psi (+0)  \cr    \psi (l-0)  \cr } , \quad
    \Psi' := \pmatrix{ \psi'(+0)  \cr  - \psi'(l-0)  \cr } ,
    \eqn\safam
    $$
where $\II$ is the two-by-two identity matrix, $L_0$ is an
auxiliary nonzero constant possessing the dimension of length [\FT,
\Ann] and $\psi' = \d\psi/\d x$.  An alternative expression is provided by 
    $$
    U \bit \Psi\PL = \Psi\MI \co \hskip 10ex
    \Psi\PM := \Psi \pm iL_0 \bit \Psi' ,
    \eqn\colbv
    $$
which we shall use later. The characteristic matrix \mm{U \in U(2) \cong U(1)
\times SU(2)} can be conveniently parametrized as
    $$
    U = e^{i \xi} \pmatrix{ \alpha & \beta \cr - \beta^* & \alpha^* \cr }
      = e^{i \xi} \pmatrix{ \aR + i \aI & \bR + i \bI \cr
      - \bR + i \bI & \aR - i \aI \cr } ,
    \eqn\stparm
    $$
with $\xi \in [0, \pi)$ and $\alpha, \beta \in \C$ satisfying
    $$
    | \alpha |^2 + | \beta |^2 = \aR^2 + \aI^2 + \bR^2 + \bI^2 = 1 \pe
    \eqn\aadl
    $$
Thus the self-adjoint Hamiltonian is characterized by the matrix $U$ and
we denote it by $H_U$.  In mathematical terms, the domain of the 
Hamiltonian is given by
$$
{\cal D}(H_U) = \big\{\psi \in {\cal H}\, \big\vert\, \psi, \, 
\psi' \in
\hbox{AC}(0, l), \,\, U\bit\Psi\PL =\Psi\MI \big\},
\eqn\domhamil
$$
where ${\cal H} = L^2(0, l)$ is the the Hilbert space
consisting of square integrable functions on the interval $(0, l)$ and
$\hbox{AC}(0, l)$ is the space of absolutely continuous functions on it.

The boundary condition (\safam) is exactly the same as the one for
the line system with a point singularity, which can be realized under some
singular potential at \mm{x = 0} [\Car --\ENZ] generalizing
the well-known Dirac delta interaction. Such a potential can be explicitly
constructed, \eg as a sequence of Dirac deltas with appropriately chosen,
diverging strengths determined by three of the four parameters of
$U$.  The remaining parameter, \mm{\arg \beta \co} which represents the phase
jump of the wave function at \mm{x = 0 \co} can be realized as a vector
potential.  On a line, this vector potential is unphysical and can be gauged
away [\JMP], since the relative phase of the two opposite sides of
the singularity cannot be measured there.  However, 
on a circle this is no longer true, that is, the phase can be measured by
interference of states because the two sides of the singularity are 
connected.  Indeed, the phase jump expresses the magnetic flux
that penetrates through the circle and is by all means physical.

Now, to study the spectral property of the circle system, we first consider 
the positive spectrum provided by the 
energy eigenfunctions of the form
    $$
    \phy_k (x) = A_k \, e^{ikx} + B_k \, e^{-ikx} \co \qquad k > 0 \pe
    \eqn\aaak
    $$ 
For such a wave function, the boundary vectors are
    $$
    \Psi = \tau_k \pmatrix{ A_k \cr B_k }, \xex{10}
    \Psi' = i k \bit \sigma_3 \tau_k \bit \sigma_3 \pmatrix{ A_k \cr B_k },
    \eqn\aaec
    $$
where \mm{\sigma_k}, \mm{ k = 1, 2, 3 } denote the Pauli matrices and
    $$
    \tau_k := \pmatrix{ 1 & 1 \cr e^{ikl} & e^{-ikl} } .
    \eqn\aaee
    $$
Then, the connection condition (\safam) reads
    $$
    [ \bit (U - \II) \tau_k - k L_0 (U + \II) \sigma_3 \tau_k \bit
    \sigma_3 \bit ] \pmatrix{ A_k \cr B_k } = 0 \co
    \eqn\aaed
    $$
or, explicitly,
    $$
    \pmatrix{ \alpha K_- + ( \beta e^{ikl} - e^{-i \xi} ) K_+ &  
    \alpha K_+ + ( \beta e^{-ikl} - e^{-i \xi} ) K_- \cr
    \alpha^* e^{ikl} K_+ - ( \beta^* + e^{-i \xi} e^{ikl} ) K_- &
    \alpha^* e^{-ikl} K_- - ( \beta^* + e^{-i \xi} e^{-ikl} ) K_+
    \cr} \pmatrix{ A_k \cr B_k \cr } = 0,
    \eqn\aaal
    $$
with $ K_\pm := 1 \pm kL_0 $. To have a nontrivial solution for the
coefficients $A_k$, $B_k$, the determinant of the matrix of the lhs of
(\aaal) must be zero. This gives the condition
    $$
    \[ \beta_{\rm I} + \sin \xi \, \cos kl \] + \left[ ( 
    \cos \xi - \alpha_{\rm R} ) + ( \cos \xi + \alpha_{\rm R}) \,
    (k L_0)^2 \right] \, \f{ \sin{kl} }{ 2 k L_0 } = 0
    \eqn\aaam
    $$
for the wave number $k$. The positive spectrum is an
infinite discrete series, in which, for large $k$, the difference
between subsequent levels is getting closer and approaches
$\pi / l$ (see Appendix A).  For the negative spectrum one only needs to replace $ik
\to \kappa$ in the formulas above to obtain the corresponding condition, 
    $$
    \[ \beta_{\rm I} + \sin \xi \, \cosh{\kappa l} \] + \[ ( \cos\xi -
    \alpha_{\rm R} ) - ( \cos \xi + \alpha_{\rm R} ) \, (\kappa L_0)^2
    \] \, \f{ \sinh{\kappa l} }{ 2 \kappa L_0 } = 0 \pe
    \eqn\aaao
    $$
{}From (\aaao) one finds that at most two negative energy states can
exist. Similarly, for a possible zero energy state, the $k \to 0$ limit
can be used to obtain the corresponding condition
    $$
    \[ \beta_{\rm I} + \sin \xi \] + \[ \cos \xi - \alpha_{\rm R} \]
    \ff{l}{ 2 L_0 } = 0 \pe
    \eqn\aaat
    $$

At this point we note that, despite that system is free on \mm{ 0
< x < l \co} negative energy states may appear because the point
singularity can act as an attractive potential, as in the special
case of a Dirac delta potential with negative strength.  A reflecting
wall, which also admits potential-type realizations [\Wall], allows
maximally one negative energy state [\Wall], while our circle system
allows maximally two as in the line system [\JMP].  We also note
that, similarly to the line system, for some special $U$ such as
\mmm{ U = - \sigma_1} the ground state may be doubly degenerate.
This is not in conflict with the well-known property of nondegeneracy
in energy levels of one dimensional quantum mechanics, because the
premises used to prove the property do not hold here.  Indeed, in
such a finite system neither the wave function nor its derivative
at infinity are required to vanish, and even at the singularity
this is not required unless our boundary condition happens to impose
it.

In fact, one can determine when an energy
eigenstate (ground state or higher) becomes doubly degenerate\note{%
Degeneracy higher than two does not arise since the energy eigenvalue equation is
a second order differential equation.} 
as follows.  Observe first that,
for states with positive energy $E > 0$,
degeneracy occurs when all the four elements of the matrix \m{U} in (\aaal)
are zero. {}From this one derives 
    $$
    \aI = \bR = 0 \co \hpt{70} \bI \ne 0 \co
    \eqn\aacv
    $$
and, further, the conditions for the energy eigenvalue
    $$
    \bI \cos kl = - \sin \xi \co
\qquad
    \bI \, k L_0 \sin kl = - ( \cos \xi - \aR ) \co \vph{\Bigg|}
\qquad 
\bI \sin kl = - ( \cos \xi + \aR ) \, k L_0
    \eqn\aads
    $$
in addition to (\aaam).  Since (\aacv) implies
$
    \aR^2 + \bI^2 = 1 \co
$
from (\aads) we find
    $$
    \xex{4.}  k^2 L_0^2 ( \cos \xi + \aR ) = \cos \xi - \aR \pe
    \eqn\aadt
    $$
On can obtain the conditions for states with \mm{E = 0} and \mm{E < 0} 
analogously, and the result is that, in both cases, one has (\aacv)
and 
    $$
    \xi = \arccot \ff{l}{2 L_0}
    \eqn\aacw
    $$
together with (\aads) with \mm{k = 0} for \mm{E = 0 \co}
or (\aads) with \mm{k \to -i \kappa} 
for \mm{E < 0}.
One then finds 
that degeneracy of an \mm{E \le 0} eigenvalue excludes any other
degeneracies, and that, unless \mm{ \cos \xi = - \aR \co } (\aadt)
can hold for only one \m{k}. If \mm{ \cos \xi = - \aR \co }
one has \mm{ \cos \xi = \aR }
(see (\aadt) and its \mm{E \le 0} variants) and, consequently,
\mm{ \xi = \pi/2 \co} \mm{ \aR = 0 } and \mm{\bI = \pm 1}. 
This shows that in the
\m{SU(2)} family there are only two types of singularities specified by 
\mm{ U = \pm \sigma_1 } that admit double degeneracy 
with more then one energy levels. 
Actually, for the cases \mm{ U = \pm \sigma_1 } 
all the positive energies prove to be doublets.  Further, 
the case \mm{U = \sigma_1}
possesses a singlet zero energy state as the ground state while 
\mm{U = - \sigma_1}
does not have any nonpositive energies.  This (almost) entire degeneracy
of energy levels suggests that the system may be bestowed supersymmetry, which
we shall confirm later.

Now we come to the point to discuss the spectral space of the circle system with a
point singularity, that is, we determine the entirety of distinct spectra
that can arise on the circle under the 
\m{U(2)} family of point interactions. {}From the spectral
conditions (\aaam)--(\aaat) we can see immediately that the spectrum
depends at most on the three parameters, \m{\xi}, \m{\aR} and \m{\bI}, of the four
of $U \in U(2)$, even though the eigenstates depend on all of the
four parameters in a nontrivial way [see (\aaal)]. 
We have also seen that the conditions
for an energy to be degenerate  depend only on
the same three parameters.  The question is thus whether these three
parameters, \m{\xi}, \m{\aR} and
\m{\bI}, really index different spectra.  This can be answered affirmatively by
a detailed examination on the possible spectra and their connection with the set of
parameters.  In fact, our argument presented in Appendix A
shows that the spectrum of a circle system uniquely
determines the parameters \m{\xi}, \m{\aR} and \m{\bI} and, consequently, the spectral
space $\Sigma_{\rm circle} := \{{\rm Spec}(H_U) \,\vert\, U \in U(2)\}$ is
given by 
$$
\Sigma_{\rm circle} = 
    \left\{ (\xi, \aR, \bI) \in \R^2 \ \big\vert  
\ \xi \in [0, \pi), \,\,\aR^2 + \bI^2 \leq 1  \right\} 
\homeo S^1 \times D^2,
\eqn\cbc    
$$
which is topologically a filled torus.
The disc $D^2$ part of $\Sigma_{\rm circle}$ may equally be realized by
$SU(2) / U(1)
\homeo S^3 / S^1$, where the \mm{SU(2)\homeo S^3} given by
    \E{\cbd}{
    \{ (\aR, \aI, \bR, \bI) \in \R^4 \ | \
    \aR^2 + \aI^2 + \bR^2 + \bI^2 = 1 \}
    }
is factorized by the phase of \mmm{ \aI + i \bR } which forms the
\mm{U(1)\homeo S^1}.  We shall encounter this latter identification later,
when we discuss a circle with two singularities.  
It is interesting to compare the spectral space (\cbc) with that of the
line system which is two dimensional and is given by the M\"obius strip
with boundary, \mm{\Sigma_{\rm line} \homeo ( S^1 \times S^1 ) / \Z_2 }
[\JMP].

\ve
\secno=3 \meqno=1

\sectit{3. Generalized symmetries, symmetries and invariant subfamilies}

It has been known that, for the line system with a point
singularity, (generalized) symmetries play an important role in classifying the
singularities [\Ann].   We shall see that
this is also the case for the circle system here.   But, first, let us recall
what we mean by a (generalized) symmetry.  Given a system with a singularity
characterized by $U$, we call a
unitary or antiunitary transformation \m{\qV} of the wave functions, 
\mm{\psi  \arrow{\qV}   \tilde{\psi}
= \qV \psi , } {\it symmetry} if it commutes with the differential operator
\m{\qHam} and further if 
\m{\tilde{\psi}} also satisfies the boundary condition (\safam) that
\m{{\psi}} fulfills, {\it i.e.}, if it commutes with $H$ including the
domain, $[\qV, H_U] = 0$.  Conversely, given a transformation \m{\qV}, one may
find, among the entire family
\m{{\cal F} = U(2)} of singularities, the subfamily
\m{{\cal F}_\qV \subset {\cal F}} which is a set of $U$ for which  \m{\qV} is a
symmetry.  Even if \m{\qV} which commutes with the differential operator
\m{\qHam}
is not a symmetry for $U \in {\cal F}$, it may still induce a map 
\mm{U \arrow{\qV} {U}_\qV \in {\cal F} }.  This motivates us to define {\it
generalized symmetry} as transformations that (are unitary or antiunitary and
commute with \m{\qHam} and) map any $U \in {\cal F}$ to another, 
generally different \mm{{U}_\qV\in {\cal F} }.  Since \m{\qV} commutes with $H$,
the generalized symmetry assures that the two systems, one characterized by $U$
and the other by \m{{U}_\qV}, share the same spectrum.

Before investigating various symmetries and generalized
symmetries arising for the circle systems, we mention a formula
valid for a certain important class of transformations and becomes convenient
in the subsequent discussions.
Suppose that a transformation $\qW$ of the wave functions, 
\mm{\psi \arrow{\qW} \tilde{\psi} =
\qW \psi
\co } commutes
with \m{\qHam} and induces transformations on the boundary vectors in (\safam)
as
    \E{\aadm}{
    \Psi \arrow{\qW} \tilde{\Psi} = M \Psi \co \xex{10} 
\Psi' \arrow{\qW} \tilde{\Psi}' = N
\Psi'
    }
with some two-by-two matrices \m{M} and \m{N}.   Then, in terms of 
\m{\Psi\PL} and \mm{ \Psi\MI = U \Psi\PL} defined in (\colbv) we have
    \E{\aadn}{
    \tilde{\Psi}\PM = M \Psi \pm i L_0 N \Psi' = \ff{1}{2} \bit
    \Bf{[} M (\II + U) \pm N (\II - U) \Bf{]} \bitt \Psi\PL \co
    }
and hence
    \E{\aado}{
    \tilde{\Psi}\MI =
    \Bf{[} M (\II + U) - N (\II - U) \Bf{]} \bitt
    \Bf{[} M (\II + U) + N (\II - U) \Bf{]}^{-1} \bitt \tilde{\Psi}\PL 
    }
as long as the inverse matrix in question exists.  We thus see that if
    \E{\aadp}{
    {U}_\qW := \Bf{[} M (\II + U) - N (\II - U) \Bf{]}
    \bittt      \Bf{[} M (\II + U) + N (\II - U) \Bf{]}^{-1} 
    }
is unitary and hence belongs to $U(2)$, then $\qW$ is a generalized symmetry. 
In particular, when \m{\,M = N \in U(2)}, which we will meet frequently
below, (\aadp) reduces to 
    \E{\aadq}{
    {U}_\qW = M \bit U \biT M^{-1} \pe
    }
Since this ${U}_\qW$ belongs to $U(2)$, such a \m{\qW} commuting with $H$ is a
generalized symmetry. If, in addition, \m{U} commutes with $M$, then
one has ${U}_\qW = U$ and hence such \m{\qW} is a symmetry.

Specializing to the circle system, the first example of symmetry transformations
we wish to mention is the parity (or space reflection), \m{\P}, defined as
    $$
    \psi(x) \arrow{\P}  ( \P \psi ) (x) = \psi (l - x) \pe
    \eqn\aadb
    $$
It clearly commutes with the Hamiltonian, and its action on the
boundary vectors [see (\safam) and (\colbv)] is found readily to be
of the form (\aadm) with \mm{M = N = \sigma_1} and, hence, the parity
\m{\P} is a generalized symmetry.  Indeed, 
\mm{{U}  \arrow{\P} {U}_\P = \sigma_1 U \sigma_1} implies
    $$
    \xi    \arrow{\P}   \xi      \co \hpt{20}
    \alpha \arrow{\P}   \alpha^* \co \hpt{20}
    \beta  \arrow{\P} - \beta^*  \co
    \eqn\aade
    $$
and thus the spectral parameters \m{\xi}, \m{\aR} and \m{\bI} remain the same, as required.  Since
\mm{\sigma_1^2 = \II}, the parity \m{\P} induces {\it duality} in spectrum in
the family ${\cal F}$ of singularities on a circle.   Note that for systems
with \m{U} satisfying $[U, \sigma_1] = 0$, the parity \m{\P} is a symmetry,
and that such a $U$ has such parameters \m{\xi, \alpha, \beta} that
\mm{\aI = 0} and \mm{\bR = 0}. The set of those $U$ form the parity invariant
subfamily $\qFP$ which, in view of (\aadl), reads
$$
\qFP \homeo S^1 \times S^1
\subset {\cal F}.
\eqn\no
$$

We may consider a one-parameter family (\m{U(1)} group) of generalized
symmetries constructed from the parity \m{\P} used
as an infinitesimal generator,
  \E{\aadv}{
  \P_\teta := e^{-i \f{\teta}{2} \P} = \cos \f{\teta}{2} \II
  - i \sin \f{\teta}{2} \P \co \xex{10} \teta \in [ 0, 2\pi ) \pe
  }
These transformations also commute with $H$ and act on
the boundary vectors as (\aadm) with \mmm{ M = N =
e^{-i \f{\teta}{2} \sigma_1} }, and are thus generalized
symmetries. Their physical effect is incorporated through the transformations
of the \m{U(2)} parameters:  
\m{\xi},
\m{\aR} and \m{\bI} are kept invariant, 
while a rotation is induced among \m{\bR} and \m{\aI} as
  \E{\aadw}{
  \bR + i \aI \bitt \arrow{\P_\teta} \bitt e^{i \teta} ( \bR + i \aI ) \pe
  }
This means that \m{\P_\teta} furnishes a rotation among the spectrally
identical point interaction systems in the parameter space, and that systems
that are invariant under
\m{\P_\teta} are those with \mmm{ \bR = \aI = 0 \co } which, to no surprise,
is the parity invariant subfamily \m{\qFP}.  Now the point is that, because of this
$U(1)$  group of generalized symmetries within the family ${\cal F} = U(2)$, 
the spectral space is found to be the coset, 
$$
\Sigma_{\rm circle} = U(2)/U(1) = U(1) \times [SU(2) / U(1)],
\eqn\cssp
$$
which is
precisely the result (\cbd).

Another important discrete transformation worth mentioning is the time
reflection,
    $$  \psi    \arrow{\T} \T \psi = \psi^* ,  \eqn\aadf  $$
which leaves $H$ invariant.  It transforms the boundary vectors as
    $$
    \Psi    \arrow{\T} \Psi^*    \co \hpt{30}
    \Psi'   \arrow{\T} {\Psi'}^* \co \hpt{30}
    \Psi\PM \arrow{\T} {\Psi\MP}^* \co
    \eqn\aadg
    $$
and, consequently, maps the characteristic matrix to its transposed,
$ U \arrow{\T} U_\T = U^T \in SU(2)$. This shows that the time
reflection $\T$ is a generalized symmetry, although it does not belong to
the class mentioned in (\aadm), being actually antiunitary.  In terms of the
parameters, we find
    $$
    \xi    \arrow{\T}   \xi     \co \hpt{20}
    \alpha \arrow{\T}   \alpha  \co \hpt{20}
    \beta  \arrow{\T} - \beta^* \co
    \eqn\aadi
    $$
and hence the spectrum is preserved. 
Clearly, $\T$ is a duality and 
the time reversal invariant subfamily \m{\qFT}
consists of those $U$ with $U = U^T$, {\it i.e.}, with 
\m{\bR = 0}, and hence
$$
\qFT \homeo S^1 \times S^2
\subset {\cal F}.
\eqn\no
$$
We also mention that
the two duality transformations, $\P$ and $\T$,  can be combined to give
the space-time reflection operator \m{\PT}.  On $U$ it acts as
$
    U \arrow{\PT} U_\PT = \sigma_1 \bitt U^T \bit \sigma_1
$
and hence
    $$
    \xi    \arrow{\PT} \xi      \co \hpt{20}
    \alpha \arrow{\PT} \alpha^* \co \hpt{20}
    \beta  \arrow{\PT} \beta    \pe
    \eqn\aadk
    $$
The subfamily \m{\qFPT} of \m{\PT}-invariant \m{U} is
determined by \mm{\aI = 0}, and hence
$$
\qFPT \homeo S^1 \times S^2
\subset {\cal F}.
\eqn\no
$$
Clearly, from neither \m{\qFP} nor \m{\qFPT} one can define a one-parameter family of
generalized symmetries analogous to $\P_\teta$.

We remark at this point that, in the line system, in addition to the
parity
\m{\P} we have two other duality transformations, \m{\qQ} and \m{\qR}, so
that the three $\{\P, \qQ, \qR\}$ form an
\m{su(2)} algebra [\Ann].  These \m{su(2)} elements can then be used as
generators to form a three-parameter (\m{SU(2)} group of) family of generalized
symmetries on the line [\JMP], as we did above for the parity $\P$ only.  This
in turn implies that, generically, the set of systems sharing the same spectrum
as a given system is larger (because the generalized
symmetries are larger in dimension) on a line, and hence the
spectral space $\Sigma_{\rm line}$ is smaller.
Heuristically speaking, the
nontrivial topology, the finite geometry and the corresponding
extra length scale cause more variety of spectra on a circle,
where \m{\qQ} and \m{\qR} are ill-defined.

\ve
\secno=4 \meqno=1

\sectit{4. Supersymmetry}

We have encountered in sect.2 two cases in the
\m{{\cal F} = U(2)} family where all the positive energy states are
doubly degenerate.  These cases are characterized by \mm{U = \pm \sigma_1}, and
we examine now if these can be interpreted as supersymmetric. 

One might think that this is trivial, since the Hamiltonian is the same
differential operator as of the free system, and hence the supercharges, 
  \E{\aadz}{
  Q_1 := \f{\h}{2i \sqrt{m}} \f{\d}{\d x} \co \xex{10} Q_2 := i \P Q_1,
  }
will clearly fulfill the algebraic relation of supersymmetry 
  \E{\aaea}{
  \{ Q_i, Q_j \} = \qHam_U^{} \bit \delta_{ij} \co \xex{10} i, j = 1, 2 \pe
  }
However, the point is that the algebra (\aaea) should hold also in the sense of
domains, not just in the differential operator relation, and it is a nontrivial
question if the domains ${\cal D}(Q_i)$ of the supercharges $Q_i$ for
$i = 1$, 2 can be given so that they can meet this demand. 

To see that this is indeed the case, we first note that, for the two cases in
question, \mm{U = \pmeps \sigma_1} with \mm{\pmeps = \pm 1}, the domains
(\domhamil) of the Hamiltonian read
  $$
  {\cal D}(H_{\pmeps \sigma_1}) = \big\{\psi \in {\cal H}\, \big\vert\, \psi,
  \, \psi' \in \hbox{AC}(0, l), \,\, \psi(l) = \pmeps \psi(0), \, \psi'(l) =
  \pmeps \psi'(0) \big\}.
  \eqn\domhamilss
  $$
Now, if we provide the domains ${\cal D}(Q_i)$ as
$$
{\cal D}(Q_i) = \big\{\psi \in {\cal H} \,\big\vert\, \psi \in
\hbox{AC}(0, l), \, \psi(l) = \pmeps \psi(0) \big\} \co
\eqn\domsc
$$
we can readily confirm that $Q_i$ are self-adjoint on these domains [\RS]. 
Moreover, by using the formulae
$$
{\cal D}(A+B) = {\cal D}(A) \cap {\cal D}(B), \qquad
{\cal D}(AB) = \{\psi \in {\cal D}(B)\, \vert\, B\psi \in {\cal D}(A)\},
\eqn\no
$$
for the domains of the sum and the product of any two linear operators $A$ and
$B$, we see immediately that the domain of the lhs of (\aaea) coincides with the
domain (\domhamilss).  We therefore conclude that the systems \mm{U = \pm
\sigma_1} indeed possess an \mmm{N = 2} supersymmetry. 

Note that for \mm{U = \sigma_1} the ground state is unique and hence
the supersymmetry is unbroken (or \lq good\rq), whereas for \mm{U = -
\sigma_1} the ground state is doubly degenerate and supersymmetry is
broken.  \note{Incidentally, we point out that here the Witten
parity operator [\Junker] is played by the parity \m{\P}.}  
Due to the topology of the circle, the possibility of supersymmetry
is limited compared to the line system where a richer
variety of supersymmetric systems have been be found [\UTone, \Kobe],
under a slightly generalized supercharges.

\ve
\secno=5 \meqno=1

\sectit{5. More subfamilies and the WKB exactness}

We have seen in sect.3 that generalized symmetries can be used to
define various subfamilies, such as \m{\qFT}, \m{\qFP}, \m{\qFPT}
as the set of the singularities for which  the respective generalized
symmetry is actually a symmetry.  There are, however, some other
subfamilies which are defined without using the generalized symmetries
and admit salient properties in the spectrum and the WKB exactness.
In this section we discuss these properties in some detail, providing
a fuller account of our earlier result in Ref.[\FT] (from which we
adopt the notations for the subfamilies).

We begin our discussion with the {\it separated} subfamily, $\qF_1 \subset {\cal
F}= U(2)$, which is the set of singularities that prohibit the probability
current $j(x) = - {{i\hbar}\over{2m}}\left(
(\psi^*)'\psi - \psi^* \psi' \right)(x)$
from flowing through the singularity, \mm{ j(l-0)
= j(+0) = 0 }.  This condition is fulfilled by diagonal 
$U$, \ie by those with \mm{\beta = 0}, and the boundary conditions
(\safam) split into two separate ones,
    $$
    \psi(l-0) + \Ll \, \psi'(l-0) = 0 \co \xex{7}
    \psi(+0) + \Lr \, \psi'(+0) = 0 \co
    \eqn\bcwell
    $$
for the left and right sides of the singularity respectively, where
\mm{ \Llr = L_0 \cot \f{\xi \pm \arccos \aR }{2} \pe}
Obviously, the cutoff of physical contact at $x = 0$ allows us to regard such a
system effectively an interval $(0, l)$, in other words, a box or infinite
potential well system. Among this subfamily $\qF_1$ are four special cases
$(\Ll, \Lr) = (0, 0)$,
$(\infty, \infty)$, $(0, \infty)$, $(\infty, 0)$, in which the theory is
explicitly solvable [\FT]. For example, the Feynman kernel is found to be
    $$ K(b, T; a, 0) = \sqrt{\ff{m}{2\pi i \hbar T}} \sum_{n
    = -\infty}^\infty \epsilon^n \left( e^{{i\over\hbar}
    {m\over{2T}}\left\{(b - a) + 2nl\right\}^2} \mp e^{{i\over\hbar}
    {m\over{2T}}\left\{(b + a) + 2nl\right\}^2} \right) \co
    \eqn\fkwallthree $$
where the \quotes{$-$}-sign is for \mm{\Ll = 0} and the \quotes{$+$}-sign
is for \mm{ \Ll = \infty }, and $\epsilon$ is $1$ for \mm{(\Ll,
\Lr) = (0, 0)} and \mm{ (\infty, \infty) }, and is $-1$ for the
two other cases. This propagator is WKB-exact in the
sense that it is a sum of free WKB amplitudes contributed by all
possible classical paths that lead from $(a, 0)$ to $(b, T)$, including those
that perform bouncing motion, hitting the left wall \mm{n} times and the
right one
\mm{n} or \mm{ n \pm 1 } times (depending on the initial direction of the
particle). Even the appearing $\pm 1$ factors allow a WKB interpretation since
one can observe that any $-1$ factor belongs to a bouncing on a
reflecting wall with \mm{L = 0} and the $1$ ones to bouncing on a
wall with \mm{L = \infty}, in view of the fact [\Wall] that, based
on some appropriate realizing potential sequences for a reflecting
wall, an \mm{L = 0} wall picks up a WKB factor \mm{-1}, while an
\mm{L = \infty} wall has the WKB factor 1.

The second subfamily we mention is the {\it scale
independent} subfamily $\qF_2$ consisting of systems for which the
coefficients $A$, $B$ in the eigenfunctions [\cf (\aaak)] are
$k$-independent. This happens for the characteristic matrices $U$
with $ \xi = \f{\pi}{2} $ and $\aR = 0$ [which form a sphere
$S^2 \subset {\cal F}$], and for $U = \pm \II$ [two isolated points in
${\cal F}$] (see Appendix B). These are the cases where the boundary
conditions do not mix the boundary values of $\psi$ with values of
$\psi'$.  More explicitly, in these cases \mm{\Ll} and \mm{\Lr} are zero and
hence the scale constant $L_0$ does not appear in the boundary conditions,
leaving $l$ as the only scale parameter.  One 
may therefore expect that, in the limit $l \to \infty$, the system becomes a
scale invariant \QPIL\ system. Indeed, it has been known 
[\Ann] that, on the line, systems belonging to the subfamily $\qF_2$
are those which are invariant under the dilatation symmetry 
\mmm{( \qW_\lambda \psi ) (x) = \lambda^{\f{1}{2}} \psi (\lambda x)}.
As for $\qF_1$, the systems belonging to $\qF_2$ can be solved [\FT],\note{%
The intersection of the subfamilies $\qF_1$ and $\qF_2$ consists
of the two special cases $ U = \pm \sigma_3 $, which are 
the box systems $(\Ll, \Lr) = (0, \infty)$, $(\infty, 0)$. Two
other important special cases in $\qF_2$ are $U = \pm \sigma_1$, which
have already been discussed in sect.3. Note that the energy eigenfunctions of $U
= \sigma_1$ provide just the basis $\{\cos nx, \, \sin nx\}$ that is used in the
classic Fourier expansion.} and the Feynman kernel can be obtained
explicitly [\FT]. For a generic $ U \in \qF_2$, using the notations
    $$ C_\pm = \f{ (1 + \aI) + (\bI - i \bR) e^{\pm i \theta}
    }{ 2 \sqrt{ (1 + \aI) (1 - \bI^2) } }\co \qquad \theta =
    \arg (\beta) \co \eqn\coeff $$
one finds
    $$ K(b, T; a, 0) = \sqrt{\ff{m}{2\pi i \hbar T}} \sum_{n =
    -\infty}^{\infty} \biggl\{ M_n e^{{i\over\hbar} {m\over{2T}}\{(b -
    a) + nl\}^2} - N_n e^{{i\over\hbar} {m\over{2T}}\{(b + a) + nl\}^2}
    \biggr\} \co \eqn\fktwo $$
with
    $$ M_n = | C_+^{} |^2 e^{-i \theta n} + | C_-^{} |^2 e^{i \theta n},
    \qquad
    N_n = C_-^* C_+^{} e^{-i \theta n} + C_+^* C_-^{} e^{i \theta n}
    \eqn\cddef $$
Unlike in the previous subfamily $\qF_1$, however, the factors
$M_n$,
$N_n$ do not admit a semiclassical interpretation, as
one can readily confirm by using (\cddef) and (\coeff) together with
\mm{|\alpha|^2 + |\beta|^2 = 1} that, \eg $|M_n| < 1$ for generic
$n$.  The situation is similar to the half line systems with a wall
that have a finite $L$, for which the bounce factors are not phase factors
[\Wall]. Consequently, we can apply the result found there, that is,
such bounce factors cannot be given
a semiclassical realization.  Hence, generically, the WKB exactness is
not perfect in the subfamily $\qF_2$.

However, there is a subfamily within the family \m{\qF_2} where
the WKB exactness holds perfectly.  It is the {\it smooth} subfamily \m{\qF_3},
containing the cases in \m{\qF_2} with \mm{\aI = 0 \pe } \m{\qF_3} is a
one-parameter U(1) subfamily, parametrized solely by the \m{\theta} of
above. The boundary conditions here read
    $$
    \psi (0) = e^{i\theta} \psi (l) \co \qquad
    \psi'(0) = e^{i\theta} \psi'(l) \co
    \eqn\aacy
    $$
which are nothing but the boundary conditions for the smooth circle
[\Schulman], \ie for the circle with no singularity.  As mentioned
in sect.2, the phase parameter \m{\theta} is regarded as the flux of a
magnetic field penetrating through the circle. In this subfamily, the propagator
(\fktwo) simplifies to the well-known kernel of the smooth circle
    $$
    K(b, T; a, 0) = \sqrt{ \ff{m}{2\pi i \hbar T}}
    \sum_{n = -\infty}^{\infty} e^{i\theta n}
    e^{{i\over\hbar} {m\over{2T}}\[(b - a) + nl\]^2} \co
    \eqn\aacz
    $$
which is readily seen to be WKB exact --- the $n$th term belongs to a
classical path on which the particle takes \m{n} turns before reaching
the point \m{b}, without acquiring any additional action contribution
each time when it crosses the point \mm{x = l \equiv 0 \pe}

Another subfamily worth mentioning is the {\it isospectral} subfamily
\m{\qF_4}, comprising those \m{U} with \mm{\xi = 0} and \mm{\bI =
0}. These systems are peculiar in that they possess the same positive
energy spectrum, \mmm{ k = n \pi / l \kern10pt ( n = 1, 2, \ldots
) \co } independently of \m{U}, although the possible
zero or negative energy is \m{U} dependent.
 This subfamily admits a generalization to the {\it semi-isospectral}
subfamily
\m{\qF_5}, characterized by the condition \mm{\sin \xi = \pm \bI
\co} where the positive spectrum consists of two infinite sequences,
one that is equidistant and \m{U}-independent and another one that
is \m{U}-dependent and given by transcendental roots of (\aaam).

\ve
\secno=6 \meqno=1

\sectit{6. A circle with two singularities}

The spectral properties and the (generalized) symmetries of a circle system
with a singularity may also be studied by considering it as a special
case of a system with two singularities.  We shall see in this section that
the latter, extended system has a graded
structure which offers a convenient framework to discuss the
spectrum preserving generalized symmetries on a direct basis, which is also
useful to study the spectral properties for the system with one singularity.

Let us place an extra singularity at $x = l/2$ on the circle, and thereby
consider the spectral properties of the modified system.  This modification
allows us to regard the system as a pair of two subsystems, one given by the
interval $[0, l/2)$ and the other by $[l/2, l)$ connected appropriately at the
ends. Given a state $\psi(x)$ on the circle, we define the corresponding state
$\Psix(x)$ by
$$
\Psix(x) = \left( {\matrix{ {\psi_+(x)}\cr {\psi_-(x)}\cr} } \right)
        := \left( {\matrix{ {\psi(x)}\cr {\psi(l - x)}\cr} }
          \right), \qquad 0 \leq x < l/2 \co
\eqn\aael
$$
on which our Hamiltonian acts by
$$
H = -{\hbar^2\over {2m}}\frac{d^2}{dx^2}\otimes \II \pe
\eqn\ha
$$

Now, let $U_1$, $U_2 \in U(2)$ be the two characteristic
matrices specifying the two singularities at $x = 0$ and $x = l/2$,
respectively. Then the connections conditions read
$$
\eqalign{
(U_1-\II)\Psix(0)+iL_0(U_1+\II)\Psix^\prime(0)&=0, \cr
(U_2-\II)\Psix(l/2)+iL_0(U_2+\II)\Psix^\prime(l/2)&=0.
}
\eqn\nyoro
$$
Note that the system is characterized by the pair $(U_1, U_2)$, and hence
there exists an ${\cal F}(U_1, U_2) = U(2) \times U(2)$ family of systems if we
allow both elements of the pair to vary in the group $U(2)$. If we freeze one
of them, for instance, by putting $U_2 = \sigma_1$ to realize the free
connection condition at $x = l/2$, then we recover the original family
${\cal F}(U_1, \sigma_1) = U(2)$ of systems with one singularity.
An important point to be noted is that, if $\Psix$ obeys (\nyoro), then the new
state
$$
\tilde \Psix := V\Psix, \qquad \hbox{for} \quad V \in SU(2),
\eqn\no
$$ 
obeys the connection conditions (\nyoro) with $U_1$, $U_2$ replaced by
$\tilde U_1$, $\tilde U_2$ which are given by the conjugation,
$$
\tilde U_1 = V U_1 V^{-1}, \qquad \tilde U_2 = V U_2 V^{-1}.
\eqn\conjpair
$$
Observe that for the Hamiltonian (\ha) we have $[V, \,H] = 0$ with $V \in SU(2)$ 
being regarded as a multiplication operator. This implies that, if
$\Psix$ [obeying (\nyoro)] is an eigenstate of $H$ with energy $E$,  so is
$\tilde \Psix = V\Psix$, that is,
$$
\hbox{if} \quad H\Psix = E\Psix, 
\qquad \hbox{then} \quad  H\tilde \Psix = E\tilde \Psix
\quad \hbox{for} \quad \tilde \Psix = V\Psix.
\eqn\isc
$$ 
This shows that the transformation $\Phi \mapsto\tilde\Phi$ is a generalized
symmetry and that there exists an isospectral family of systems which are
related by (\conjpair).  

Now we specialize to the case with one singularity at $x = 0$ by choosing $U_2 =
\sigma_1$. The isospectral family is then obtained by collecting those pairs
$(\tilde U_1,
\tilde U_2)$ in (\conjpair) for which $\tilde U_2 = U_2 = \sigma_1$.  Obviously,
the solution is given by $V = e^{i\alpha \sigma_1}$, whose
totality provides the isospectral group $U(1)$. 
Namely, if we let ${\cal I}(U_1, U_2)$ be the isospectral group for the system
specified by the pair $(U_1, U_2)$, then we have
${\cal I}(U_1, \sigma_1) = U(1)$.
Accordingly, the spectral space for the circle system with two singularities,
$$
\Sigma(U_1, U_2) := {\cal F}(U_1, U_2)/{\cal I}(U_1, U_2)
\eqn\no
$$ 
is given by
$$
\Sigma(U_1, \sigma_1) = U(2)/U(1),
\eqn\no
$$
which is in fact the spectral space $\Sigma_{\rm circle}$ (\cssp).

Another special case arises when the second singularity does not allow the
probability current flow through it and is also parity invariant.
This system is actually an interval of length $l$ with a singularity at $x = 0$
with walls having an identical boundary condition for the right and the left
sides, and this arises when $U_2$ belongs to the {\it self-dual} [\duallet, \Ann]
subfamily $ \qFSD \subset \qFP$, \ie taking the form
\mm{U_2 = e^{i\theta} \II \co \xex{2} \theta \in [0, 2\pi )}.  In this case,  the
isospectral family given by those pairs of $(\tilde U_1, \tilde U_2)$ for which
$\tilde U_2 = U_2 = e^{i\theta} \II$ is just the entire $SU(2)$, that is,
${\cal I}(U_1, e^{i\theta} \II) = SU(2)$. Thus the spectral space is found to be
$$
\Sigma(U_1, e^{i\theta} \II) = U(2)/SU(2)  \homeo S^1 \times S^1/\Z_2,
\eqn\no
$$
which is just $\Sigma_{\rm line}$ we know of [\JMP].

The spectral space of a system with a generic (but fixed) $U_2$ is also 
studied analogously.  For this, one just decomposes $U_1$ and $U_2$ as
$$
U_i = V_i^{-1} D_i V_i, \qquad
V_i \in SU(2), \qquad
D_i =
    \left( \matrix{    
        e^{i\theta_i^+} & 0 \cr
        0 & e^{i\theta_i^-}
    } \right), \qquad i = 1,2.
\eqn\decomp
$$
Then, one finds from the
isospectral conjugation that its spectral space is identical to that of the
diagonal $U_2 = D_2$ (since it is obtained by choosing $V = V_2$ in (\isc)). 
For
$U_2 = D_2$, the isospectral group is determined by those $V \in SU(2)$ for
which $VD_2V^{-1} = D_2$.  If $\theta_2^+ = \theta_2^-$, such $D_2$ belongs to
the self-dual subfamily and the isospectral group ${\cal I}(U_1, D_2)$
is given by $SU(2)$ as mentioned above.  
If $\theta_2^+ \neq \theta_2^-$, the isospectral group
${\cal I}(U_1, D_2)$ is the $U(1)$ consisting of the elements $V = e^{i\rho
\sigma_3}$.  Clearly, the generic $U_2 = V_2^{-1} D_2 V_2$ has the
isospectral group 
${\cal I}(U_1, U_2)$ given by the $U(1)$ consisting of the elements
$V = V_2^{-1} e^{i\rho\sigma_3} V_2 = e^{i\rho V_2^{-1}\sigma_3 V_2}$.
Thus, for generic $U_2 = V_2^{-1} D_2 V_2$ where $V_2$ and $D_2$ are fixed with
$\theta_2^+ \neq \theta_2^-$, we find
${\cal I}(U_1, V_2^{-1} D_2 V_2) = U(1)$
and
$$
\Sigma(U_1, V_2^{-1} D_2 V_2) = U(2)/U(1).
\eqn\no
$$

The foregoing discussions allude us to ask the total spectral space that
arises if we are allowed to vary both $U_1$ and $U_2$, that is, in circle
systems with the total family ${\cal F}(U_1, U_2) = U(2) \times U(2)$.
Due to the isospectral
transformation (\isc), we have  the isospectral group,
${\cal I}(U_1, U_2) = SU(2)$,
and hence the total spectral space is expected to be
$$
\Sigma(U_1, U_2) = [U(2) \times U(2)]/SU(2).
\eqn\no
$$
We need, of course, to supplement our argument to prove this by showing that
each element of the spectral space leads to a distinct spectrum, as has been
done in sect.3 (with Appendix A) for the case of one singularity.

We close our discussion by mentioning briefly
the generalizations of the connection formulas to the present case of 
two singularities.
{}For this we introduce the combined boundary vectors
    $$
    \Psi  := \pmatrix{ \Psix (0) \cr \Psix (l/2) } , \xex{10}
    \Psi' := \pmatrix{ \Psix'(0) \cr \Psix'(l/2) } ,
    \eqn\aaef
    $$
and the combined characteristic matrix
    $$
    U := \pmatrix{ U_1 & 0 \cr 0 & U_2 } ,
    \eqn\aaeg
    $$
and thereby rewrite the connection conditions (\nyoro) into an analogous 
form of (\safam) as
    $$
    (U - \IIII) \Psi + i L_0 (U + \IIII) \Psi' = 0,
    \eqn\aaeh
    $$
where $\IIII$ is the four-by-four unit matrix. 
{}For the energy eigenvalue problem, the positive energy eigenfunctions
are of the form
    $$
    \Psix(x) = \pmatrix{ A_k \, e^{ikx} + B_k \, e^{-ikx} \cr
                         C_k \, e^{ikx} + D_k \, e^{-ikx} } ,
    \eqn\aaei
    $$
with boundary vectors
    $$
    \Psi  = T_k \pmatrix{ A_k \cr B_k \cr C_k \cr D_k } , \xex{10}
    \Psi' = T_k \, \Sigma_3 \pmatrix{ A_k \cr B_k \cr C_k \cr D_k } ,
    \eqn\aaej
    $$
where
    $$
    T_k = \pmatrix{ 1 & 1 & 0 & 0 \cr 0 & 0 & 1 & 1 \cr
    e^{ikl/2} & e^{-ikl/2} & 0 & 0 \cr  0 & 0 & e^{ikl/2} & e^{-ikl/2} }
    , \xex{6} \Sigma_3 = \pmatrix{ 1 & 0 & 0 & 0 \cr 0 & -1 & 0 &  0 \cr 
                                   0 & 0 & 1 & 0 \cr 0 &  0 & 0 & -1 } .
    \eqn\aaek
    $$
The boundary conditions therefore read
    $$
    [ \bit (U - \IIII) \, T_k - k L_0 (U + \IIII) \, T_k \, \Sigma_3 \bit ] 
    \pmatrix{ A_k \cr B_k \cr C_k \cr D_k } = 0 \pe
    \eqn\aaem
    $$
These formulas are the generalizations of (\aaec)--(\aaed). The
slight difference that only $T_k \bit \Sigma_3$ is needed instead of $
\Sigma_3 \bit T_k \bit \Sigma_3$ is a consequence of the convenient definition
of the second component of $\Psix$ [see (\aael)]. The nonpositive energy
eigenfunctions can be obtained from the positive energy ones the same way as
in sect.3.  

The special cases we discussed above may then be explicitly examined
by choosing the matrix $U$ appropriately.  In particular, for the
spectral properties, one can use the isospectral transformation
(\isc) to specify $U$ further and thereby simplify the discussion
considerably [\UTone, \UTtwo].  {}For instance, for the generic
case, (\isc) can be used to diagonalize either $U_1$ or $U_2$ in
determining the spectrum.  This unified matrix form of the boundary
conditions furnishes a useful tool for more than two singularities
and, generally, for any larger set of boundary conditions, too.

\ve
\secno=7 \meqno=1

\sectit{7. Conclusions}

We have studied in this paper the spectral properties of a quantum
particle on a circle with a pointlike singularity and found that,
when considering the whole four-parameter family of possible point
singularities, the space of distinct spectra forms a three-dimensional
set given by $\Sigma_{\rm circle} \homeo S^1 \times D^2$.  This
result is contrasted to the case of point singularities on the
line, where the spectral space is two dimensional and given by
$\Sigma_{\rm line} \homeo ( S^1 \times S^1 ) / \Z_2$.  By analyzing
the possible generalized symmetries of the circle systems, we have
found that the difference between the two spectral spaces can be
attributed to the difference in the symmetry structure --- the
structure of the line systems is somewhat richer than that of the
circle system.

Further, we have determined when the circle systems possess degenerate
energy levels, and proved that the cases where all the positive
energy levels are doubly degenerate actually exhibit an $N = 2$
supersymmetry.   The class of supercharges considered is a standard
one to realize the supersymmetry algebra, and we have shown that
the algebra holds including the domains for the operators involved.
The question of domain in the supersymmetry algebra is a nontrivial
mathematical problem and has been answered only partially for simple
systems such as lines/intervals [\UTone, \UTtwo].  On the physical
aspects, on the other hand, the dependence of the degeneracy of an
energy eigenstate on the parameters characterizing the point
singularity opens the possibility that, if a quantum nanodevice
realizing a point interaction on a circle can be fabricated with
tunable parameters, then experiments demonstrating coherent control
of quantum states will be possible to carry out (experiments like,
\eg the ones reported in [\NPT]).

We have also extended earlier results on the connection between
symmetries and special subfamilies of the four-parameter full family
of point singularities, where  the subfamilies are defined with
respect to the symmetries such as parity and time reflection.  In
a few special subfamilies, {\it i.e.}, separated subfamily or smooth
subfamily, there are cases where the WKB exactness is observed in
the transition amplitude.  This implies that in those cases the
amplitude admits the interpretation that it is  a sum of contributions
of free propagations plus a certain bouncing/penetration effect
occurring at the singularity.

Finally, we have put our discussion on the generalized symmetries
with one point singularity in the context of two singularities on
a circle.  There, we have seen that both of the spectral spaces,
$\Sigma_{\rm circle}$ and $\Sigma_{\rm line}$, appear as special
cases, showing that this provides a scheme to discuss the spectral
properties of these systems on a general basis.  Obviously, this
generalization should be useful to incorporate further singularities
allowing for their classification by symmetries.  Increasing the
number of singularities is not just a matter of mathematical
extension, because repeated structures such as lattices do appear
in physics even in one dimension, and we hope that our methods and
results presented in this paper can be extended to those systems
which undoubtedly have richer and more interesting properties.

\bigskip
\noindent
{\bf Acknowledgements:}

This work has been supported in part by the Grant-in-Aid for
Scientific Research on Priority Areas (No.~13135206) and (C)
(No.~13640413) by the Japanese Ministry of Education, Science,
Sports and Culture.

\ve

\secno=0 \appno=1 \meqno=1

\sectit{Appendix A. The spectral space $\Sigma_{\rm circle}$}

To prove that a spectrum uniquely determines \m{\xi}, \m{\aR} and
\m{\bI}, first we study the dependence of the spectrum on these
parameters. When doing this, it is useful to recall that \mm{\xi
\in [0, \pi) \co} and \mmm{\aR^2 + \bI^2 \leq 1 \pe}

To start with, if \mmm{\xi = \bI = 0} (let us call this case `case I') 
then the positive energies satisfy
    \E{\cad}{
    [ ( 1 - \aR ) + ( 1 + \aR ) (k L_0)^2 ] \sin kl = 0 \pe
    }
Since \mm{|\aR| \leq 1}, \mm{1 - \aR} and \mm{1 + \aR} are 
non-negative, and only one of them can be zero, therefore 
\mmm{ [ ( 1 - \aR ) + ( 1 + \aR ) (k L_0)^2 ] } is always positive. 
Consequently, the positive energies fulfil \mm{\sin kl = 0}, with the 
solutions \mmm{ k_n = \f{\pi}{l} n \co \ \ n = 1, 2, \ldots }

If at least one of \m{\xi} and \m{\bI} is nonzero then let us first
consider the subcase \mmm{\aR = - \cos \xi} (called `case II'). Here,
\mmm{\sin \xi \neq 0}, since \mmm{ \sin \xi = 0 \impl \xi = 0 \impl \aR =
-1 \impl \bI = 0 \co } and this case is now excluded. Thus we can write
(\aaam) in the form
    \E{\cat}{
    \ff{\bI}{\sin \xi} + \cos kl + \ff{\cot \xi}{k L_0} \sin kl = 0 \pe
    }
On the left hand side, if \mmm{k \to \infty} then the third term tends 
to zero so \mmm{\f{\bI}{\sin \xi} + \cos kl \to 0 \co} \mmm{\cos kl \to 
- \f{\bI}{\sin \xi} \pe} This means that the large roots \m{k} will get 
closer and closer to the values of the form \mmm{ \f{1}{l} ( \arccos
\f{- \bI}{\sin \xi} + 2 \pi n ) \co } respectively, \mmm{ \f{1}{l} ( -
\arccos \f{- \bI}{\sin \xi} + 2 \pi n ) \co } in an alternating sequence.

In the remaining case --- \ie when \mmm{\aR \neq - \cos \xi} and at 
least one of \m{\xi} and \m{\bI} is nonzero `case III') --- (\aaam) can 
be written as
    \E{\caf}{
    \f{a_1}{kl} + \f{a_2}{kl} \cos kl +
    \left[ \f{a_3}{(kl)^2} + 1 \right] \sin kl = 0 \co
    }
with
    \E{\cam}{
    a_1 = \ff{ 2 \bI }{ \cos \xi + \aR } \ff{l}{L_0} \co \xex{6}
    a_2 = \ff{ 2 \sin \xi }{ \cos \xi + \aR } \ff{l}{L_0} \co \xex{6}
    a_3 = \ff{ \cos \xi - \aR }{ \cos \xi + \aR } \( \ff{l}{L_0} \)^2 \pe
    }
If \mmm{k \to \infty} then the terms proportional to \m{a_1}, \m{a_2} 
and \m{a_3} tend to zero so \mmm{\sin kl \to 0 \pe} Now the large roots 
\m{k} are getting closer and closer to the values of the form \mmm{ 
\f{\pi}{l} n \co \ \ n \in \Z \pe } 

Now we will determine more details about the asymptotic behaviour of the 
roots \m{k}. For this reason, we make the following Ansatz: 
    \E{\cag}{
    k_n l = \pi n + \eps_n \co \qquad \hbox{where} \qquad
    \eps_n = \ff{c_1}{n} + \ff{c_2}{n^2} + \ff{c_3}{n^3} + \cdots \pe
    }
The coefficients \m{c_i} can be determined iteratively from (\caf). For 
our purposes the first three coefficients will be the interesting ones. 
To be up to this order, it is enough to use the formulae
    \E{\cah}{
    \cos k_n l = \cos (\pi n) \cos \eps_n =
    (-1)^n [ 1 - \ff{1}{2} \eps_n^2 ] + \ordo{\eps_n^4} \co
    }
    \E{\cai}{
    \sin k_n l = \cos (\pi n) \sin \eps_n = 
    (-1)^n [ \eps_n - \ff{1}{6} \eps_n^3 ] + \ordo{\eps_n^5} \co
    }
    \E{\caj}{
    (k_n l)^2 = \pi^2 n^2 + 2 \pi c_1 + \ordo{ \ff{1}{n} } \co
    }
    \E{\cak}{
    \eps_n^2 = \ff{c_1^2}{n^2} + \ordo{ \ff{1}{n^3} } \xex{10}
    \eps_n^3 = \ff{c_1^3}{n^3} + \ordo{ \ff{1}{n^4} } \pe
    }
Inserting these into the condition (\caf) multiplied, for convenience, 
by \mm{(kl)^2 \co} and grouping the terms as decreasing powers of \m{n}, 
the vanishing of the coefficients of \m{n^1}, \m{n^0} and \m{n^{-1}} 
lead to
    \E{\can}{
    c_1 = - \ff{1}{\pi} [ (-1)^n a_1 + a_2 ] \co
    \xex{6} c_2 = 0 \co \xex{6} c_3 =
    \ff{- c_1}{\pi^2} a_3 + \ff{c_1^2}{6 \pi} (c_1 + 3 a_2 - 6) \co
    }
respectively. We can see that there is a sequence \mmm{ c_1^{(+)}, 
c_2^{(+)}, c_3^{(+)}, \ldots } for even \m{n}s, and another sequence 
\mmm{ c_1^{(-)}, c_2^{(-)}, c_3^{(-)}, \ldots } for odd \m{n}s. Note 
that at least one of \m{a_1} and \m{a_2} is nonzero, because \mmm{\xi = 
\bI = 0} is now excluded. Therefore, at least one of \m{ c_1^{(+)} } and 
\m{ c_1^{(-)} } is nonzero. Thus in case III the roots do not 
\emp{exactly} fulfil \mmm{\sin kl = 0}, they are only getting closer and 
closer to it, \mm{\sin kl} only \emp{tends to} zero.

In the possession of this collected knowledge, we can turn to the 
inverse problem we wish to solve, \ie to identify the parameters \m{\xi},
\m{\aR} and \m{\bI} from a given spectrum.

If all the positive energies satisfy \mmm{\sin kl = 0} \emp{exactly} 
then we can know that we are in case I. This determines \m{\xi} and 
\m{\bI} (namely, \mmm{\xi = \bI = 0}) but \m{\aR} is yet unknown. Let us 
see whether the possible zero and negative energies determine \m{\aR}. 
The condition for a zero energy state (\aaat) reads in this case simply 
\mm{\aR = 1 \pe} Therefore, if the spectrum contains a zero energy state 
then \mm{\aR = 1 \pe} If not, then let us see the possibility for
negative energies: (\aaao) is now
    \E{\caq}{
    [ (1 - \aR) - (1 + \aR) (\kappa L_0)^2 ] \sinh \kappa l = 0 \co
    }
which gives that there exists one negative energy state with
    \E{\car}{
    \kappa = \ff{1}{L_0} \sqrt{ \ff{1 - \aR}{1 + \aR} }
    }
if \mm{\aR \neq -1} and no negative energy state if \mm{\aR = -1 \pe} 
Consequently, from the absence of negative energy states we learn 
\mm{\aR = -1}, and from one negative energy state with \m{\kappa} we 
can identify \m{\aR} as
    \E{\cas}{
    \aR = \ff{ 1 - (\kappa L_0)^2 }{ 1 + (\kappa L_0)^2 } \pe
    }

If we see that \mm{\cos kl} tends to a definite value as \m{k} increases
then we know that we face at case II [since in case I \mm{\cos kl}
oscillates between 1 and \m{-1}, and in case III \mmm{ \cos ( k\_{n = 2j}
l ) \to 1 } and \mm{ \cos ( k\_{n = 2j + 1} l ) \to -1 } ]. {}From \mmm{
\lim ( \cos kl ) } we obtain \mm{ \bI / \sin \xi \co } and then, from
(\cat), using any root \m{k} from the known spectrum for which \mmm{\sin
kl \neq 0 \co} we determine \mm{\cot \xi}, which uniquely tells \m{\xi}.

In the end, if we find that the positive spectrum is such that the 
values of \mm{\sin kl} tend to zero but are not exactly zero then we 
know we are in case III. For large enough \m{k}s we can determine which 
integer \m{n} belongs to a \m{k} (by rounding \mm{ kl / \pi } to the 
nearest integer). Then, we can identify the coefficients
\mm{ c_1^{(+)} } and \mm{ c_3^{(+)} } as
    \E{\cau}{
    c_1^{(+)} = \lim_{ \scriptstyle n \to \infty \atop \scriptstyle
    n \, {\rm even} } n ( k_n l - \pi n ) \co \xex{8}
    c_3^{(+)} = \lim_{ \scriptstyle n \to \infty, \atop \scriptstyle
    n \, {\rm even} } n^3 \[ k_n l - \pi n - \ff{ c_1^{(+)} }{n} \] ,
    }
and \mm{ c_1^{(-)} } and \mm{ c_3^{(-)} } in a similar way. {}From \mm{ 
c_1^{(+)} } and \mm{ c_1^{(-)} } we can obtain \mm{a_1} and \mm{a_2} 
[\cf (\can)] as
    \E{\caw}{
    a_1 = - \ff{\pi}{2} \[ c_1^{(+)} - c_1^{(-)} \] \co \qquad \qquad
    a_2 = - \ff{\pi}{2} \[ c_1^{(+)} + c_1^{(-)} \] \co
    }
and then, corresponding to that which of \mm{ c_1^{(+)} } and \mm{ 
c_1^{(-)} } is nonzero --- we know that at least one of them is 
nonzero ---, \mm{a_3} can be determined from \mm{ c_3^{(+)} } or
\mm{ c_3^{(-)} \co } respectively [\cf (\can)].

{}From \mm{a_1}, \mm{a_2} and \mm{a_3} the parameters \mm{\xi}, \mm{\aR} 
and \mm{\bI} are calculated as follows [all steps will be based on
(\cam)]. If \mmm{a_3 = - (l/L_0)^2} then \mmm{ \cos \xi = 0 \quad \impl
\quad \xi = \pi / 2 }, and \mmm{ \aR = 2 / [ (L_0 / l) a_2 ] } and \mmm{ 
\bI = a_1 / a_2 \pe } If \mmm{a_3 \neq - (l/L_0)^2} then, observing that 
    \E{\caz}{
    \ff{ (L_0 / l) \bittt a_2 }{ 1 + (L_0 / l)^2 \bit a_3 } = \tan \xi,
    }
\m{\xi} is determined uniquely. Then, we have
    \E{\cba}{
    \aR = \ff{ 1 - (L_0 / l)^2 a_3 }{ 1 + (L_0 / l)^2 a_3 } \cos \xi
    \co \xex{10} \bI = \ff{a_1}{a_2} \sin \xi \pe
    }

We can summarize the above considerations with that the spectrum of a 
circle system uniquely determines its parameters \m{\xi}, \m{\aR} and 
\m{\bI}.

\medskip
\secno=0 \appno=2 \meqno=1

\sectit{Appendix B. The scale independent boundary conditions}

On dimensional grounds, the coefficients $A$, $B$ in the eigenfunctions
(\aaak) will be $k$-independent if \m{L_0} actually drops out from
the boundary conditions expressed by (\safam). This happens if both
lines of the matrix equation (\safam) --- or, two appropriate linear
combinations of them --- contain only one of the two boundary value
vectors \m{\Psi}, \m{\Psi'}.

First, suppose that neither of the rows of the matrices \mm{U - \II} and
\mm{U + \II} are identically zero. Then an appropriate linear combination of
the two lines of (\safam) is needed to drop \m{\Psi} out from, say, the
first line. In this case, any other linear combination will leave some
\m{\Psi} in the second line so the goal of another linear combination
will be to drop \m{\Psi'} out from the second line. This is possible
only if both matrices \mm{U - \II} and \mm{U + \II} are such that their
first row is a multiple of their second row. Then we have
    $$
    \det (U - \II) = \det (U + \II) = 0 \co
    \eqn\aada
    $$
which tells that the two eigenvalues of \m{U} are \m{\pm 1}.
Therefore, \mm{ U = P_+ - P_- }, where \m{P_+} is the projector projecting
onto the eigensubspace of \m{U} corresponding to the eigenvalue 1, and
\m{P_-} projecting onto the other eigensubspace. {}From this we see that
\m{U} is self-adjoint, and this property leads to the requirements \mm{
\xi = \f{\pi}{2} }, \mm{ \aR = 0 \pe }

Second, if some of the rows of \mm{U - \II} and \mm{U + \II} are identically
zero then first we observe that this can happen to at most two of the four
rows in question: Otherwise at least one of the matrices \mm{U - \II \co}
\mm{U + \II} would be zero, but then the other one should be \mm{\pm 2 \II},
which has only nonzero rows. Now, if two rows of the four are zero then
it is easy to see that one of these rows must be an upper row and the
other a lower row (the difference of \mm{U + \II} and \mm{U - \II} is \m{2 \II},
which makes the other cases impossible). This means four possibilities,
the matrices \mm{ U = \pmatrix{ \pm 1 & 0 \cr 0 & \pm 1 } \co } which
give the two isolated scale independent systems \mm{ U = \pm \II } (the
two other cases, \mm{U = \pm \sigma_3 \co} are included in the \m{U}s
with \mm{ \xi = \f{\pi}{2} }, \mm{ \aR = 0 }). At last, if one of the
four rows is zero --- say, a row of \mm{U - \II} ---, then one of the two
lines of (\safam) is already \m{\Psi} independent. The other line will
then necessarily contain \m{\Psi} so, to make it \m{\Psi'} independent,
a suitable multiple of the \m{\Psi} independent line has to be added to
it. This means that the two rows of \mm{U + \II} has to be each other's
multiple, consequently, \mmm{ \det (U + \II) = 0 \pe } However, \mm{U - \II}
has a zero row so \mmm{ \det (U - \II) = 0 \co } too. Thus we arrive again
back to (\aada), and hence to \mm{ \xi = \f{\pi}{2} }, \mm{ \aR = 0 }.


\baselineskip= 15.5pt plus 1pt minus 1pt
\parskip=5pt plus 1pt minus 1pt
\tolerance 8000
\vfill\eject\immediate\closeout\reffile
\centerline{{\bf References}}\bigskip\frenchspacing%
\input refs.tmp\vfill\eject\nonfrenchspacing

\bye